%% file: main.tex
\documentclass[sigconf,nonacm]{acmart}

\newif\ifDEBUG
\DEBUGtrue
\DEBUGfalse

\newif\ifEXTENDED
\EXTENDEDtrue
\EXTENDEDfalse

\usepackage{graphicx}
\usepackage{hyperref} 
\usepackage{cleveref}
\usepackage{caption}
\usepackage{pifont}
\usepackage{xcolor}
\usepackage{listings}
\usepackage{minted}
\usepackage{subcaption}
\usepackage{caption}

\usepackage{booktabs}
\usepackage{multirow} 
\usepackage{subcaption}
\usepackage{makecell}
\usepackage{tabularx}
\usepackage{ragged2e}
\usepackage{xspace}
\usepackage{soul}
\usepackage{array} 
\usepackage{float}

\usepackage{seqsplit}
\usepackage{multicol}
\usepackage{placeins}
\usepackage{amsmath}
\usepackage{textcomp}
\usepackage{colortbl}
\usepackage{longtable}
\usepackage{array}
\usepackage{adjustbox}
\usepackage{algorithm}
\usepackage{threeparttable}
\usepackage{tablefootnote}
\input{misc/typesetting}
\begin{document}

\title{The Research Software Scope: what and where}
\title{Toward a Usable Taxonomy of Research Software for Supply Chain Security.}
\title{Taxonomy-Aware Measurement for Research Software Supply Chains}
\title{A Usable Taxonomy of Research Software for Supply Chain Security}
\title{A Taxonomy of Research Software for Supply Chain Security}
\title{Operationalizing Research Software for Supply Chain Security}

\author{Kelechi G. Kalu}
\affiliation{%
  \institution{Purdue University}
  \city{West Lafayette}
  \state{Indiana}
  \country{USA}
}

\author{Soham Rattan}
\affiliation{%
  \institution{Purdue University}
  \city{West Lafayette}
  \state{Indiana}
  \country{USA}
}

\author{Taylor R. Schorlemmer}
\affiliation{%
  \institution{Purdue University}
  \city{West Lafayette}
  \state{Indiana}
  \country{USA}
}

\author{George K. Thiruvathukal}
\affiliation{%
  \institution{Loyola University}
  \city{Chicago}
  \state{Illinois}
  \country{USA}
}

\author{Jeffrey C. Carver}
\affiliation{%
  \institution{University of Alabama}
  \city{Auburn}
  \state{Alabama}
  \country{USA}
}

\author{James C. Davis}
\affiliation{%
  \institution{Purdue University}
  \city{West Lafayette}
  \state{Indiana}
  \country{USA}
}

\renewcommand{\shortauthors}{Kalu et al.}

\begin{abstract}

Empirical studies of research software are hard to compare because the literature operationalizes ``research software'' inconsistently.
Motivated by the research software supply chain (RSSC) and its security risks, we introduce an RSSC-oriented taxonomy that makes scope and operational boundaries explicit for empirical research software security studies.

We conduct a targeted scoping review of recent repository mining and dataset construction studies, extracting each work's definition, inclusion criteria, unit of analysis, and identification heuristics.
We synthesize these into a harmonized taxonomy and a mapping that translates prior approaches into shared taxonomy dimensions.
We operationalize the taxonomy on a large community-curated corpus from the Research Software Encyclopedia (RSE), producing an annotated dataset, a labeling codebook, and a reproducible labeling pipeline.
Finally, we apply OpenSSF Scorecard as a preliminary security analysis to show how repository-centric security signals differ across taxonomy-defined clusters and why taxonomy-aware stratification is necessary for interpreting RSSC security measurements.

\end{abstract}




\maketitle

\section{Introduction}

Research software (\textbf{RS}) underpins modern science and engineering.
Like other software, RS depends on upstream artifacts and actors, which forms a software supply chain (SSC) that shapes how software is developed, built, distributed, and maintained~\cite{Okafor_Schorlemmer_Torres-Arias_Davis_2022}.
We refer to this dependency network for RS as the research software supply chain (\textbf{RSSC}).
Software supply chains are a frequent vector for cyberattacks~\cite{willett2023lessons,codecov_security_update_2021,eslint2018postmortem,jiang2025confuguard, kalu2026armsvisionactorreputation}, and supply chain risk is widely observed in practice~\cite{synopsys_ossra_2023}.
These risks motivate empirical RSSC security research and policy-relevant guidance grounded in evidence.

However, empirical work lacks a consistent basis for what counts as research software~\cite{gruenpeter2021defining_research_software}.
Repository mining studies and dataset papers operationalize RS differently~\cite{Murphy_Brady_Shamim_Rahman_2020,Brown_Schwartz_Huang_Weber_2024,Thakur_Milewicz_Jahanshahi_Paganini_Vasilescu_Mockus_2025, chuehong2022fair4rs}.
These differences include scope boundaries,
units of analysis (for example, repositories versus packages),
and identification heuristics.
For example, some studies link repositories to publications~\cite{park2019research,orduna2021link}, some use funding linkage~\cite{Brown_Schwartz_Huang_Weber_2024}, and others use text-based classification of repository metadata or README content~\cite{Thakur_Milewicz_Jahanshahi_Paganini_Vasilescu_Mockus_2025}.
Such studies sample different software for their measurements, and despite using similar terminology, are in disagreement about the underlying construct.
This limits inter-study comparability and hampers cumulative RSSC security conclusions.

We address this gap by developing an RSSC-oriented taxonomy that makes operational boundaries explicit for repository-mining studies.
We derive an initial taxonomy from Okafor \etal's SSC framing of actors, operations, and artifacts and their associated security properties~\cite{Okafor_Schorlemmer_Torres-Arias_Davis_2022}, then validate and refine it through a targeted scoping review that extracts how prior work defines and operationalizes RS.
We operationalize the taxonomy on the Research Software Encyclopedia (RSE) corpus, producing an annotated dataset, a labeling codebook, and a reproducible labeling pipeline.
We then apply OpenSSF Scorecard~\cite{ossf_scorecard_website} as a preliminary security analysis to illustrate how repository-centric security signals differ across taxonomy-defined clusters, and we contextualize these signals using an enterprise-backed open source baseline from the Apache Software Foundation \cite{apache_projects_foundation_projects_json, github_apache_org}.

In summary, our contributions are as follows:
\begin{itemize}
    \item We present an RSSC-oriented taxonomy and a mapping that harmonize divergent operational boundaries used in prior repository-mining studies.
    The taxonomy makes scope choices explicit in terms of actors, operations, and artifacts, enabling clearer comparison across RS datasets and methods.
    \item We operationalize the taxonomy on the RSE corpus and release an annotated dataset, a labeling codebook, and a reproducible labeling pipeline to support reuse and replication.
    \item We apply OpenSSF Scorecard as a preliminary, taxonomy-aware security analysis and contextualize results with an Apache baseline, illustrating why explicit scope and stratification matter for interpreting RSSC security measurements.
\end{itemize}


\section{Background \& Related Work}
\label{sec:background}

This section summarizes how prior empirical studies identify research software, and explains why an RSSC-oriented taxonomy is necessary for comparable security analysis.

\subsection{Research Software Operationalization in Empirical Studies}
\label{sec:background-rs-empirical}

We use \emph{research software} (RS) to refer to software produced or adapted to support research activities and outcomes~\cite{gruenpeter2021defining_research_software}.
Although the term is widely used, empirical studies operationalize RS in inconsistent ways~\cite{gruenpeter2021defining_research_software, Sochat_May_Cosden_Martinez-Ortiz_Bartholomew_2022}.
These differences include whether research-support tooling and infrastructure are in scope alongside result-producing applications\cite{chuehong2022fair4rs}, and they directly affect which projects are included in RS datasets.

A growing body of empirical work studies RS by mining repositories or constructing RS datasets~\cite{Murphy_Brady_Shamim_Rahman_2020,Brown_Schwartz_Huang_Weber_2024,Thakur_Milewicz_Jahanshahi_Paganini_Vasilescu_Mockus_2025}.
These studies typically start from a candidate pool of software projects and apply identification heuristics such as curated lists, links to scholarly publications~\cite{park2019research,orduna2021link}, funding linkage~\cite{Brown_Schwartz_Huang_Weber_2024}, or text-based classification of repository metadata or README content~\cite{Thakur_Milewicz_Jahanshahi_Paganini_Vasilescu_Mockus_2025}.
These strategies are useful, but they often function as proxies for RS, such as treating scientific software or funding-linked software as the target population, which can narrow scope and exclude other RS categories.
Because these choices differ across papers, two studies can use the term ``research software'' while sampling different underlying populations.
Our work differs by synthesizing these operationalizations into an RSSC-oriented taxonomy and mapping that make scope choices explicit and support comparability across studies.

\subsection{From SSC to RSSC, and Why a Taxonomy is Necessary}
\label{sec:rssc-motivation}

Software rarely exists in isolation.
Instead, software depends on upstream artifacts and actors, forming a software supply chain (SSC) that influences how software is developed, built, distributed, and maintained~\cite{Okafor_Schorlemmer_Torres-Arias_Davis_2022}.
We use \emph{research software supply chain} (RSSC) to refer to the corresponding network of dependencies among actors, operations, and software artifacts that produce, distribute, and sustain RS \cite{Okafor_Schorlemmer_Torres-Arias_Davis_2022}.
Because software supply chains are a frequent vector for compromise\cite{Kalu_Singla_Okafor_Torres-Arias_Davis_2025, williams2025research}, the RSSC inherits these risks while creating additional stakes for research settings, such as threats to integrity and trust in research outputs.

Despite growing academic~\cite{nasem2025_assessing_research_security_efforts},
governmental~\cite{whitehouse_nspm33_2021, jason2023researchsecurity},
and adversarial~\cite{gimbals2025_foreign_threats_academia_national_security} interest in RS and its security implications, the literature lacks a shared taxonomy designed for RSSC security research.
Supply chain security approaches distinguish entities such as actors, operations, and artifacts and motivate security properties that depend on what is considered in scope~\cite{Okafor_Schorlemmer_Torres-Arias_Davis_2022}.
However, existing datasets~\cite{Murphy_Brady_Shamim_Rahman_2020, Brown_Schwartz_Huang_Weber_2024, Thakur_Milewicz_Jahanshahi_Paganini_Vasilescu_Mockus_2025} and heuristics~\cite{Sochat_May_Cosden_Martinez-Ortiz_Bartholomew_2022} often embed implicit assumptions about what counts as RS, what unit is analyzed, and what evidence is sufficient for inclusion.
These choices differ across papers and communities, limiting evidence-building and translation of empirical findings into actionable RSSC guidance.

Taxonomic confusion has consequences.
Studies that claim to analyze RS may analyze different underlying populations.
Security conclusions can be non-comparable across papers, even when they use similar terminology.
Policy guidance may unintentionally target only narrow slices of the ecosystem and may not be effective for the broader RSSC.
\textit{These limitations motivate our goal of developing an RSSC-oriented taxonomy that makes scope and operationalization explicit and supports comparable security measurement.}

\section{Methodology}
\label{sec:method}

\begin{figure*} 
    \centering
    \includegraphics[width=0.68
    \linewidth]{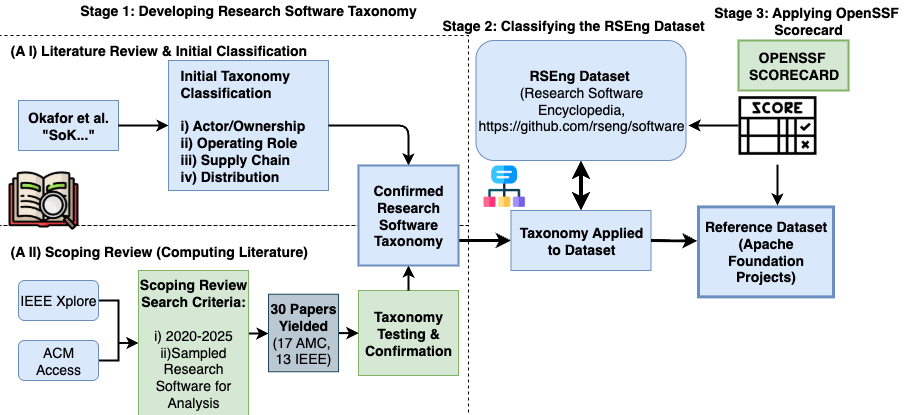}
    \caption{
        Methodology overview illustrating the taxonomy development and measurement with OpenSSF Scorecard. \GKT{Check OpenSSF throughout for consistent capitalization. I fixed it here..}
    }
    \label{fig:methodology-overview}
\end{figure*}

Our methodology follows the staged process shown in \cref{fig:methodology-overview}.
In Stage~1, we develop and validate an RSSC-oriented taxonomy that captures how prior empirical studies operationalize research software.
In Stage~2, we apply this taxonomy to label the RSE corpus.
In Stage~3, we use these labels to structure a preliminary, taxonomy-aware security measurement with OpenSSF Scorecard, contextualized with an Apache Software Foundation baseline.

\subsection{Stage 1: A research software taxonomy}
\label{sec:method-stage1}

Stage~1 combines an RSSC security framing with evidence from empirical research software studies.
We first derive candidate taxonomy dimensions from Okafor \etal's SSC conceptualization of actors, operations, and artifacts and the security properties that motivate what must be visible and verifiable in a supply chain~\cite{Okafor_Schorlemmer_Torres-Arias_Davis_2022}.
We then validate and refine these dimensions through a targeted scoping review of recent empirical studies that identify research software via repository mining or dataset construction.

\textbf{Initial taxonomy:}
We define 4 taxonomy classes that capture supply-chain-relevant distinctions: who maintains software, how it is used, and how it is delivered.
\emph{Actor unit} is the primary organizational unit responsible for producing, maintaining, or operating the software~\cite{Okafor_Schorlemmer_Torres-Arias_Davis_2022}.
\emph{Supply chain role} describes where an artifact sits in the RSSC and how compromise would propagate.
\emph{Research role} captures whether software directly produces research results or primarily supports the research process.
\emph{Distribution pathway} refers to how software is delivered for downstream use.

\textbf{Scoping review and refinement:}
We search IEEE Xplore and the ACM Digital Library from 2020 through 2025.
We screen abstracts and include papers whose abstracts indicate mining software repositories of research or scientific software projects.
For each included paper, we extract the study's operationalization of research software, including any stated definition, inclusion and exclusion criteria, unit of analysis, and identification heuristics.
We map these extractions to the initial taxonomy and refine labels and decision rules when recurring patterns are not captured.
We summarize the resulting taxonomy in \cref{tab:taxonomy-categories}.
Search queries and extraction details are in the appendix.

\subsection{Stage 2: Classifying the RSEng dataset}
\label{sec:method-stage2}

In Stage~2, we apply the validated taxonomy to a large community-curated corpus from the Research Software Encyclopedia (RSE).
For each RSE entry, we collect repository metadata and the repository \texttt{README.md} as the primary context for classification.
We then label each entry using the taxonomy codebook and an LLM pipeline.

\textbf{LLM-assisted labeling and quality checks:}
We use OpenAI GPT-5.1 to label the full corpus.
We select GPT-5.1 after small-sample checks that compare model outputs across versions and against manual labels, and after confirming that disagreements concentrate in the most interpretive boundary cases.

\subsection{Stage 3: Taxonomy-aware security measurement with OpenSSF Scorecard}
\label{sec:method-stage3}

In Stage~3, we connect taxonomy to measurement by applying OpenSSF Scorecard to the taxonomy-labeled corpus~\cite{ossf_scorecard_website}.
Our goal is not to provide a definitive assessment of RSSC security, but to show through a scalable preliminary analysis, why explicit scope and taxonomy-aware stratification matter for interpreting repository-centric security signals.
Scorecard computes signals that reflect repository-visible development and release practices, such as branch protection, code review, and CI configuration~\cite{ossf_scorecard_website,zahan2023openssf_snp}.
We analyze Scorecard results within taxonomy-defined clusters, such as dependency versus application artifacts, package registry versus other distribution pathways, and individual versus community-maintained projects.
We treat Scorecard outcomes as partial indicators, since repository-centric signals do not capture off-repository operations.
This limitation reinforces the need for a taxonomy that makes boundary choices explicit.
We additionally run Scorecard on a baseline set of enterprise-backed open source projects from the Apache Software Foundation.
We use this baseline only to contextualize the magnitude and variability of Scorecard signals under a well-resourced governance model, not to rank research software against Apache projects.

\section{Results}
We present the result of our taxonomy efforts on the RSE corpus.
To illustrate the result for JAWS, we compare the RSE corpus's security measurement (OpenSSF Scorecard)  to a baseline, both in aggregate and partitioned by actor unit.

\label{sec:results}

\subsection{Taxonomy overview on the RSE corpus}
\label{sec:results-taxonomy}

We apply our RSSC-oriented taxonomy to 6,966 research software entries from the RSE corpus.
The actor unit distribution is concentrated in three categories: research groups or labs (28.8\%), individual maintainers (28.4\%), and institutions such as universities, labs, or government research organizations (25.6\%), together covering 82.8\% of labeled entries; community or foundation-governed projects account for a further 10.0\%.
Across supply chain role, most entries are application software (65.2\%), with a substantial minority that are dependency artifacts (30.0\%); all other roles appear rarely (each under 1.3\%).
Across research role, most entries support direct research execution (63.3\%) or research-support tooling (32.9\%).
Across distribution pathway, nearly half of entries distribute via package registries (49.1\%), followed by source repositories (24.1\%) and installers or binaries (14.8\%); network services (5.0\%), containers (2.1\%), and build-and-release pipelines (1.6\%) make up the remainder.

\begin{table}[t]
\centering
\caption{
Top taxonomy categories in the RSE corpus.
Percentages are computed over all taxonomy-labeled entries.
}
\small
\begin{tabular}{p{0.22\linewidth}p{0.72\linewidth}}
\hline
\textbf{Dimension} & \textbf{Top categories} (\% of corpus) \\
\hline
Actor unit & Research group or lab (\textit{28.8\%}); Individual maintainer (\textit{28.4\%}); Institution (university, lab, government research organization) (\textit{25.6\%}); Community or foundation (open source governance) (\textit{10.0\%}) \\
Supply chain role & Application software (\textit{65.2\%}); Dependency software artifact (\textit{30.0\%}); Infrastructure (\textit{1.2\%}); Unknown (\textit{3.4\%}) \\
Research role & Direct research execution (\textit{63.3\%}); Research-support tooling (\textit{32.9\%}); Incidental or general-purpose (\textit{0.3\%}); Unknown (\textit{3.5\%})  \\
Distribution pathway & Package registry (\textit{49.1\%}); Source repo (\textit{24.1\%}); Installer or binary (\textit{14.8\%}); Network service (\textit{5.0\%}) \\
\hline
\end{tabular}
\label{tab:taxonomy-overview}
\end{table}

\subsection{Taxonomy-aware Scorecard summary with ASF baseline}
\label{sec:results-scorecard-asf}

\input{summary_table.tex}

\subsubsection{Overall comparison against the ASF baseline}
\label{sec:results-scorecard-asf-overall}

OpenSSF Scorecard produces an overall score by aggregating a set of repository-centric checks (e.g., branch protection, CI practices, security policy, dependency pinning).
In our data, each check returns a score in $\{0,\dots,10\}$ when it can be evaluated, and a score of $-1$ when the check is not applicable or cannot be evaluated from available evidence (e.g., ``no workflows found,'' ``no pull request found,'' or ``no releases found'').
We therefore report two complementary summaries: the overall Scorecard score, and a missingness rate defined as the per-repository fraction of checks scored $-1$.
This distinction matters because a score of $0$ indicates a failed evaluated practice, while $-1$ indicates the check could not be assessed and should not be interpreted as an explicit failure.

We successfully compute OpenSSF Scorecard for 5{,}937 of 6{,}966 taxonomy-labeled RSE entries (85.2\% coverage).
\cref{tab: summary-scorecard} compares research software (RS) to a baseline set of repositories owned by the Apache Software Foundation (ASF).
We summarize both overall score and missingness using medians and IQRs to reduce sensitivity to skew and long tails common in repository metrics.

Overall, RS repositories have a lower median Scorecard score than ASF (2.9 vs 3.9; $\Delta=-1.0$).
However, median missingness is similar in magnitude between RS and ASF (22\% vs 28\%), indicating that the lower RS score is not explained solely by a higher fraction of unevaluable checks.
Instead, the gap is consistent with RS repositories meeting fewer of the evaluated repository practices captured by Scorecard, on average.

\subsubsection{Taxonomy-aware comparisons by actor unit}
\label{sec:results-scorecard-asf-actor}

\Cref{tab: summary-scorecard} also illustrates how taxonomy-aware stratification changes the interpretation of Scorecard.
In this table, ASF is treated as a fixed baseline in every row: the first row compares all RS to all ASF, while subsequent rows compare RS subsets against the same ASF baseline.
The final column ($\Delta$ median) is the median RS score minus the median ASF score, so negative values indicate the RS subset scores $<$ ASF.

Actor stratification shows that differences are not uniform across RS.
For example, RS projects maintained by a community or foundation are closest to the ASF baseline (median 3.6; $\Delta=-0.3$) and also exhibit lower missingness (11\% vs ASF's 28\%), consistent with more standardized repository workflows and observable governance signals.
In contrast, RS maintained by individual maintainers shows a larger score gap relative to ASF (median 2.7; $\Delta=-1.2$) with missingness comparable to ASF in the median case (28\%).
This pattern suggests that the difference is less about Scorecard being unable to evaluate these repositories and more about the corresponding practices (e.g., branch protection, code review, dependency management automation) being less frequently adopted or less consistently configured.
Finally, the small ``Unknown'' actor category has both the lowest median score (2.5) and the highest missingness (33\%), consistent with a heterogeneous set of repositories where repository-centric signals are less consistently observable.

For completeness, we caution that actor rows with small sample sizes (e.g., mixed responsibility) should be interpreted descriptively rather than as stable estimates.
Taken together, these results show that treating research software as a single population can obscure meaningful differences in governance and workflow maturity that are better captured when conditioning on actor structure.





\section{Threats to Validity}
\label{sec:Threats}

\textbf{Internal threats.}
Our targeted scoping review was conducted ad hoc by a single author, so relevant studies may have been missed and screening decisions may reflect subjective judgment.
Some RSE entries could not be labeled or scored due to GitHub token failures and other tool errors, which could bias results if missingness correlates with repository characteristics, although overall coverage remained high enough that we do not expect these failures to change the high-level patterns.

\textbf{External threats.}
Our results are grounded in the RSE corpus and an Apache Software Foundation baseline, so conclusions may not generalize to closed-source research software or to research software ecosystems not represented in these corpora.
Our review currently treats only works from 2020--2025. 

\textbf{Construct threats.}
Taxonomy labels rely on \texttt{README.md} and repository metadata, so incomplete documentation can lead to ambiguous boundary cases and occasional misclassification.
OpenSSF Scorecard measures repository-visible practices and cannot capture off-repository operations, private workflows, or organizational controls, so Scorecard outcomes are partial indicators rather than a complete assessment of RSSC security posture.

\section{Next steps}

Our end goal is to enable cumulative, comparable evidence about research software supply chain (RSSC) security by making scope and operational boundaries explicit.
\textbf{Expanding data source and constructs:} while this study focuses on conventional open-source security signals, many RSSC projects developed using other methods, \eg institutional controls. We will identify and extend the taxonomy to incorporate additional evidence sources, incorporating human-factors input from the research software engineering community.
\textbf{Stratified studies:} we will use the taxonomy as a sampling and stratification mechanism, conditioning prior analyses to see the effect on previous results.
\textbf{Operational refinement:} we will refine the labeling pipeline through targeted manual audits and community feedback. 

These steps advance our goal: developing taxonomic infrastructure for security in the research software supply chain.

\textbf{Data availability:}
  A data artifact will be shared with our JAWS presentation \cite{purdue_duality_lab_cross}.

\section*{Acknowledgments}
We acknowledge support from the National Science Foundation (NSF) under Award Nos.~2537308, 2537309, and 2537310.

\bibliographystyle{ACM-Reference-Format}
\bibliography{references/reference, references/new_papers}

\appendix



\section{Search sources and time window.}
We query IEEE Xplore and ACM Digital Library.
We restrict the search window to 2020-2025 to focus on recent empirical practices.

\noindent
\subsection{IEEE EXPLORE Query}
We search for papers using research software and scientific software terms in titles and abstracts.
For IEEE Xplore, we use a query over the document title and abstract that includes the phrases, ``research software'',
``scientific software'', and open-source variants. We also select the conference filter --
 "(( ("Document Title":"scientific software" OR "Abstract": "scientific software") OR ("Document Title":"research software" OR "Abstract":"research software") OR ("Document Title":"research open-source software" OR "Abstract": "research open-source software") OR ("Document Title": "research open source software" OR "Abstract": "research open source software") OR ("Document Title": "scientific open-source software" OR "Abstract":"scientific open-source software") OR ("Document Title":"scientific open source software" OR "Abstract":"scientific open source software") OR ("Document Title":"open-source research software" OR "Abstract":"open-source research software") OR ("Document Title":"open source research software" OR "Abstract":"open source research software") OR ("Document Title":"open-source scientific software" OR "Abstract":"open-source scientific software") OR ("Document Title":"open source scientific software" OR "Abstract":"open source scientific software") ) )".

 \myparagraph{Final List of IEEE Explore Papers}
 \noindent
 \begin{enumerate}[leftmargin=*, itemsep=0pt]
\item LibRA: A Scientific Software Library of Radio Astronomy Algorithms. \url{10.23919/USNC-URSINRSM66067.2025.10907007}.
\item Insights into Optimizing Research Software: A Case of an Architecture Smell Detection Tool.\ \url{10.1109/SCAM67354.2025.00011}.
\item Scientific Software Engineering: Mining Repositories to gain insights into BACARDI.\ \url{10.1109/AERO47225.2020.9172261}.
\item How do Software Engineering Researchers Use GitHub? An Empirical Study of Artifacts \& Impact\ \url{10.1109/SCAM63643.2024.00021}.
\item An Empirical Study of High Performance Computing (HPC) Performance Bugs.\ \url{10.1109/MSR59073.2023.00037}.
\item Revisiting Impedance Spectroscopy: A Didactic Virtual Instrument for Modeling and Analysing Nanomaterials in Gas Sensors.\ \url{10.1109/SBMicro66945.2025.11197761}.
\item Technical Debt in the Peer-Review Documentation of R Packages: a rOpenSci Case Study.\ \url{10.1109/MSR52588.2021.00032}.
\item GNSS active control point determinations. \ \sloppy
\url{10.1109/ECTI-CON51831.2021.9454699}
\item Half-Precision Scalar Support in Kokkos and Kokkos Kernels: An Engineering Study and Experience Report.\ \url{10.1109/eScience55777.2022.00095}.
\item Research Configuration of Engineering Modeling Platform. \url{10.1109/SACI49304.2020.9118812}.
\item Physical Experiment Simulation Framework Based on Matlab.\ \url{10.1109/ICDCOT61034.2024.10515721}.
\item On the Developers' Attitude Towards CRAN Checks.\ \url{10.1145/3524610.3528389}.
\item Large Language Models: The Next Frontier for Variable Discovery within Metamorphic Testing?.\ \url{10.1109/SANER56733.2023.00070}.

\end{enumerate}

\subsection{ACM Digital Library Query}
For our search on the ACM Digital Library, we use an analogous query (to the one presentedd above) over title and abstract.
We provide the full database query -- "Title:("scientific software" OR "research software" OR "research open-source software" OR "research open source software" OR "scientific open-source software" OR "scientific open source software" OR "open-source research software" OR "open source research software" OR "open-source scientific software" OR "open source scientific software") OR Abstract:("scientific software" OR "research software" OR "research open-source software" OR "research open source software" OR "scientific open-source software" OR "scientific open source software" OR "open-source research software" OR "open source research software" OR "open-source scientific software" OR "open source scientific software")
"

\myparagraph{Final List of ACM Digital Library Papers}
\noindent
\begin{enumerate}[leftmargin=*, itemsep=0pt]
\item Soft-Search: Two Datasets to Study the Identification and Production of Research Software.\ \url{10.1109/JCDL57899.2023.00040}.
\item Evaluation of the Nvidia Grace Superchip in the HPE/Cray XD Isambard 3 supercomputer.\ \url{10.1145/3757348.3757359}.
\item Pricing Python parallelism: a dynamic language cost model for heterogeneous platforms.\ \url{10.1145/3426422.3426979}.
\item Executable Science: Research Software Engineering Practices for Replicating Neuroscience Findings.\ \url{10.1145/3736731.3746147}.
\item On the developers’ attitude towards CRAN checks.\ \url{10.1145/3524610.3528389}.
\item Practical Runtime Instrumentation of Software Languages: The Case of SciHook.\ \url{10.1145/3623476.3623531}.
\item Finding Metamorphic Relations for Scientific Software.\ \url{10.1109/ICSE-Companion52605.2021.00118}.
\item A curated dataset of security defects in scientific software projects.\ \url{10.1145/3384217.3384218}.
\item Contextual Understanding and Improvement of Metamorphic Testing in Scientific Software Development.\ \url{10.1145/3475716.3484188}.
\item A method and experiment to evaluate deep neural networks as test oracles for scientific software.\ \url{10.1145/3524481.3527232}.
\item AI Assistants to Enhance and Exploit the PETSc Knowledge Base.\ \url{10.1145/3750720.3757281}.
\item Collaboration Challenges and Opportunities in Developing Scientific Open-Source Software Ecosystem: A Case Study on Astropy.\ \url{10.1145/3757462}.
\item A Containerization Framework for Bioinformatics Software to Advance Scalability, Portability, and Maintainability.\ \url{10.1145/3584371.3612948}.
\item Scientific Open-Source Software Is Less Likely to Become Abandoned Than One Might Think! Lessons from Curating a Catalog of Maintained Scientific Software.\ \url{10.1145/3729369}.
\item Overcoming Barriers in Scaling Computing Education Research Programming Tools: A Developer’s Perspective.\ \url{10.1145/3632620.3671113}.
\item Bridging Disciplinary Gaps in Climate Research through Programming Accessibility and Interdisciplinary Collaboration.\ \url{10.1145/3759536.3763804}.
\item Do Current Language Models Support Code Intelligence for R Programming Language?\ \url{10.1145/3735635}.

\end{enumerate}

\section{Taxonomy Definitions and Scoping Review  Confirmation }
\label{sec: appendix-taxonomy-def}

As outlined in \cref{sec:method-stage1}, we conduct a targeted scoping review of empirical studies that construct datasets via repository mining of research or scientific software.
We extract how each included study operationalizes research software, and we use these operationalizations to validate and refine the initial taxonomy dimensions and decision rules.
Our goal is not an exhaustive, systematic literature review.
Our goal is to identify and systematize the operationalizations used in practice by recent repository-mining studies.

See~\cref{tab:taxonomy-categories} for the resulting table.

\FloatBarrier
\input{taxonomy_table_4dims.tex}
\FloatBarrier

\end{document}

%% file: misc/typesetting.tex
\usepackage{booktabs} 
\usepackage{amsmath}
\usepackage{algorithm}
\usepackage{algpseudocode}
\usepackage[normalem]{ulem} 
\usepackage{xspace}
\usepackage{relsize}
\usepackage{multirow} 
\usepackage{balance} 
\usepackage{fancyvrb} 
\usepackage{caption}
\usepackage{pifont}

\usepackage{enumitem}
\setlist[itemize]{leftmargin=*,noitemsep,topsep=0pt}
\setlist[enumerate]{leftmargin=*}

\usepackage{etoolbox}
\makeatletter
\patchcmd{\@makecaption}
	{\scshape}
	{}
	{}
	{}
\makeatletter
\patchcmd{\@makecaption}
	{\\}
	{.\ }
	{}
	{}
\makeatother

\newcommand{\ie}{\textit{i.e.,}\xspace}
\newcommand{\eg}{\textit{e.g.,}\xspace}
\newcommand{\etal}{\textit{et al.}\xspace}

\usepackage{mdframed}
\mdfsetup{skipabove=0.5\topskip,skipbelow=0.5\topskip,align=center}

\usepackage{amsthm}
\newtheorem{thm}{Theorem}

\DeclareMathSymbol{\mlq}{\mathord}{operators}{``}
\DeclareMathSymbol{\mrq}{\mathord}{operators}{`'}

\newif\ifSAVESPACE
\SAVESPACEfalse

\ifSAVESPACE

\else

\fi

\usepackage{todonotes}
\usepackage{soul}
\usepackage{fontawesome}
\usepackage{ragged2e}

\ifDEBUG
    \newcommand{\AH}[1]{\todo[color=cyan,inline]{AH:#1}}
    \newcommand{\AM}[1]{\todo[color=red,inline]{Machiry:#1}}
    \newcommand{\JD}[1]{\todo[color=yellow,inline]{JD:#1}}
    \newcommand{\SA}[1]{\todo[color=green,inline]{SA:#1}}
    \newcommand{\PA}[1]{\todo[color=orange,inline]{PA:#1}}
    \newcommand{\TS}[1]{\todo[color=lime,inline]{TS:#1}}
    \newcommand{\KR}[1]{\todo[color=yellow,inline]{Kyle:#1}}
    \newcommand{\LS}[1]{\todo[color=green,inline]{LS:#1}}
    \newcommand{\HP}[1]{\todo[color=cyan,inline]{HP:#1}}
    \newcommand{\NJE}[1]{\todo[color=red,inline]{NJE: #1}}
    \newcommand{\GKT}[1]{\todo[color=red,inline]{GKT:#1}}
    
    \newcommand{\RH}[1]{\todo[color=red,inline]{RH:#1}}
    \newcommand{\WJ}[1]{\todo[color=SkyBlue,inline]{Wenxin:#1}} 
    \newcommand{\KC}[1]{\todo[color=orange,inline]{Kelechi Says:#1}}
    \newcommand{\AG}[1]{\todo[color=orange,inline]{AG:#1}}
    \newcommand{\PJ}[1]{\todo[color=lime,inline]{PJ:#1}}
    \newcommand{\AZ}[1]{\todo[color=teal,inline]{Antonio:#1}}
    \newcommand{\PT}[1]{\todo[color=pink,inline]{Parth:#1}}
    
\else
    \newcommand{\AH}[1]{}
    \newcommand{\AM}[1]{}
    \newcommand{\JD}[1]{}
    \newcommand{\SA}[1]{}
    \newcommand{\PA}[1]{}
    \newcommand{\TS}[1]{}
    \newcommand{\KR}[1]{}
    \newcommand{\LS}[1]{}
    \newcommand{\HP}[1]{}
    \newcommand{\NJE}[1]{}
    \newcommand{\GKT}[1]{}
    \newcommand{\KC}[1]{}
    \newcommand{\RH}[1]{}
    \newcommand{\WJ}[1]{}
    \newcommand{\AG}[1]{}
    \newcommand{\PJ}[1]{}
    \newcommand{\PT}[1]{}
    \newcommand{\AZ}[1]{}
    
\fi

\usepackage{hyperref}

\usepackage{cleveref}
\crefformat{section}{\S#2#1#3}
\crefname{figure}{Figure}{Figures}
\crefname{table}{Table}{Tables}
\crefname{theorem}{Theorem}{Theorems}
\crefname{thm}{Theorem}{Theorems}
\crefname{lemma}{Lemma}{Lemmata}
\crefname{equation}{Eqt.}{Eqts.}
\crefformat{Grammar}{Grammar #1}
\crefname{appendix}{Appendix}{Appendices}
\crefname{listing}{Listing}{Listings}

\newcommand{\myparagraph}[1]{\paragraph{#1}}
\renewcommand{\myparagraph}[1]{\vspace{0.25em} \noindent \underline{\textit{#1:}}}

\usepackage{url}

\usepackage{tcolorbox}

\makeatletter
\newcommand{\linebreakand}{%
  \end{@IEEEauthorhalign}
  \hfill\mbox{}\par
  \mbox{}\hfill\begin{@IEEEauthorhalign}
}
\makeatother

\usepackage{pifont}

 \usepackage{pifont}

%% file: summary_table.tex
\begin{table*}[t]
\centering
\caption{OpenSSF Scorecard: Apache Software Foundation (ASF) vs. research software (RS), overall and by RS actor type.}
\small
\setlength{\tabcolsep}{3.5pt}
\renewcommand{\arraystretch}{1.15}
\label{tab: summary-scorecard}
\begin{tabular}{>{\raggedright\arraybackslash}p{0.33\linewidth}rrr|rrrr}
\toprule
\makecell[l]{\textbf{Group}} &
\makecell{\textbf{ASF}\\N} &
\makecell{\textbf{ASF score}\\(IQR)} &
\makecell{\textbf{ASF missing}\\(IQR)} &
\makecell{\textbf{RS}\\N} &
\makecell{\textbf{RS score}\\(IQR)} &
\makecell{\textbf{RS missing}\\(IQR)} &
\makecell{\textbf{$\Delta$}\\median} \\
\midrule
RS: All projects &
2575 & 3.9 (3.3--5.1) & 28\% (11--33) &
5937 & 2.9 (2.5--3.6) & 22\% (11--33) & -1.00 \\

\midrule

RS: Individual maintainer &
2575 & 3.9 (3.3--5.1) & 28\% (11--33) &
1778 & 2.7 (2.3--3.2) & 28\% (11--33) & -1.20 \\

RS: Research group or lab &
2575 & 3.9 (3.3--5.1) & 28\% (11--33) &
1679 & 3.1 (2.6--3.8) & 17\% (11--28) & -0.80 \\

\makecell[l]{RS: Institution (university, government, etc.)} &
2575 & 3.9 (3.3--5.1) & 28\% (11--33) &
1552 & 2.8 (2.5--3.5) & 28\% (11--33) & -1.10 \\

\makecell[l]{RS: Community or foundation (OSS governance)} &
2575 & 3.9 (3.3--5.1) & 28\% (11--33) &
531 & 3.6 (3.2--4.5) & 11\% (11--17) & -0.30 \\

RS: Vendor or commercial entity &
2575 & 3.9 (3.3--5.1) & 28\% (11--33) &
291 & 3.0 (2.6--3.8) & 17\% (11--28) & -0.90 \\

RS: Unknown &
2575 & 3.9 (3.3--5.1) & 28\% (11--33) &
99 & 2.5 (2.2--2.7) & 33\% (28--33) & -1.40 \\

RS: Mixed or shared responsibility &
2575 & 3.9 (3.3--5.1) & 28\% (11--33) &
7 & 3.0 (2.5--3.8) & 17\% (8--22) & -0.90 \\
\bottomrule
\end{tabular}

\vspace{2pt}
\parbox{\linewidth}{\footnotesize\emph{Notes:} Scorecard check scores of -1 indicate missing or not-applicable (insufficient evidence), while 0 indicates an evaluated check that failed. Missingness is the per-repository fraction of checks scored -1 and is summarized by median and IQR.}
\end{table*}

%% file: taxonomy_table_4dims.tex
\begin{table*}[t]
\centering
\small
\setlength{\tabcolsep}{4pt}
\caption{
Taxonomy categories used to summarize how empirical studies operationalize research software for RSSC analysis.
For each category, we provide a brief definition and representative papers from our scoping set.
}
\label{tab:taxonomy-categories}
\begin{tabular}{p{0.14\linewidth}p{0.16\linewidth}p{0.50\linewidth}p{0.16\linewidth}}
\hline
Dimension & Category & Definition & Papers (count) \\
\hline
Actor unit & Individual maintainer & Single person (or informal very small group) is the primary maintainer; no formal governance. & \cite{Sochat_May_Cosden_Martinez-Ortiz_Bartholomew_2022}\\
 & Research group or lab & A lab or grant/project team maintains the software (e.g., PI, students, staff). & \\
 & Institution & A formal institution (university, national lab, government research org) develops or operates the artifact. & \cite{Sochat_May_Cosden_Martinez-Ortiz_Bartholomew_2022} \\
 & Community or foundation & A broader OSS community or foundation provides governance and stewardship beyond a single lab/institution. & \cite{Sochat_May_Cosden_Martinez-Ortiz_Bartholomew_2022} \\
 & Vendor or commercial entity & A company is the primary producer or operator (commercial service, proprietary or dual-licensed tool). & \cite{Sochat_May_Cosden_Martinez-Ortiz_Bartholomew_2022}\\
 & Platform operator & An entity operating shared hosting or distribution infrastructure (e.g., registry, repo host, CI platform). & \cite{Sochat_May_Cosden_Martinez-Ortiz_Bartholomew_2022}\\
 & Mixed or shared responsibility & Responsibility is explicitly split across multiple actor units (e.g., lab produces; foundation governs releases). & \\
 & Unknown & Insufficient information to identify the responsible actor unit. & \\
\hline
Supply chain role & build and release & Tooling or infrastructure used to build, test, package, containerize, sign, or publish software artifacts (e.g., CI/CD, build pipelines). & \cite{Tam_Chua_Gallagher_Omene_Okun_DiFranzo_Chen_2023a}; \cite{Murphy_Brady_Shamim_Rahman_2020}; +14 more (17) \\
& dependency artifact & Software primarily consumed as a dependency by other software (libraries, packages, frameworks). & \cite{Zhao_Fard_2025}; \cite{Harvey_Milewicz_Trott_Berger-Vergiat_Rajamanickam_2022}; +1 more (3) \\
& distribution and governance & Infrastructure or processes for distributing, curating, cataloging, or governing software ecosystems (registries, catalogs, foundations). & \cite{Kumar_Ie_Vidoni_2022}; \cite{Brown_Schwartz_Huang_Weber_2024}; +3 more (5) \\
& unknown & Insufficient information in the paper snippet to assign a role. & \cite{Tsigkanos_Rani_Müller_Kehrer_2023}; \cite{Horváth_2020}; +2 more (4) \\
\hline
Distribution pathway & containers & Distributed primarily as container images (e.g., Docker) for deployment or reproducibility. & \cite{Tam_Chua_Gallagher_Omene_Okun_DiFranzo_Chen_2023a} (1) \\
& installer/binary & Distributed as downloadable binaries or installers. & \cite{Urlea_Urlea_Vanderbauwhede_Voinea_Nabi_2025}; \cite{Peres_Galeazzo_da_Silva-Dantas_Lisboa_de-Souza-Junior_2025} (2) \\
& network service & Delivered as a hosted service or API rather than a packaged artifact. & \cite{Green_Alam_McIntosh-Smith_Gilham_Wishart_2025a}; \cite{von-Kurnatowski_Stoffers_Weigel_Meinel_Wasser_Rack_Fiedler_2020}; +2 more (4) \\
& package registry & Distributed via a language package registry (\eg CRAN, PyPI, Maven, npm). & \cite{Kumar_Ie_Vidoni_2022}; \cite{Codabux_Vidoni_Fard_2021} (2) \\
& releases & Distributed via tagged releases and published release artifacts. & \cite{Murphy_Brady_Shamim_Rahman_2020}; \cite{Brown_Schwartz_Huang_Weber_2024} (2) \\
& source repo & Distributed primarily via a source code repository (cloned or downloaded). & \cite{Sun_Patil_Li_Guo_Zhou_2025}; \cite{Zhao_Fard_2025}; +4 more (6) \\
& unknown & Insufficient information in the paper snippet to determine distribution. & \cite{Smith_Zhang_Zhang_McInnes_Keceli_Vasan_Balay_Isaac_Chen_Vishwanath_2025}; \cite{Peng_Kanewala_Niu_2021}; +9 more (12) \\
\hline
Research role & Use in research workflow & Software used directly to produce, validate, replicate, or visualize research outcomes. & \cite{Peres_Galeazzo_da_Silva-Dantas_Lisboa_de-Souza-Junior_2025} (1) \\
 & Software as research object & Software itself is the object of study (e.g., mining, quality, longevity, practices, or ecosystem dynamics). & \cite{Brown_Schwartz_Huang_Weber_2024,Codabux_Vidoni_Fard_2021}; +9 more (11) \\
 & Foundation for research & Reusable foundations and infrastructure for research (e.g., libraries, platforms, build/distribution/dev tooling). & \cite{Green_Alam_McIntosh-Smith_Gilham_Wishart_2025a,Harvey_Milewicz_Trott_Berger-Vergiat_Rajamanickam_2022}; +4 more (6) \\
\hline
\end{tabular}
\end{table*}